\documentclass[letterpaper, conference]{IEEEtran}
\IEEEoverridecommandlockouts
\usepackage[left=1.91cm,right=1.91cm,top=2.54cm,bottom=2cm]{geometry}
\usepackage{cite}
\usepackage{amsmath,amssymb,amsfonts}
\usepackage{algorithmic}
\usepackage{graphicx}
\usepackage{textcomp}
\usepackage{xcolor}
\def\BibTeX{{\rm B\kern-.05em{\sc i\kern-.025em b}\kern-.08em
    T\kern-.1667em\lower.7ex\hbox{E}\kern-.125emX}}

\begin{document}

\title{Parameter Estimation-Based Observer for Linear Systems with Polynomial Overparametrization\\
\thanks{Financial support is in part provided by the Grants Council of the President of the Russian Federation (MD-1787.2022.4).}
}

\author{\IEEEauthorblockN{1\textsuperscript{st} Anton Glushchenko}
\IEEEauthorblockA{\textit{Laboratory No. 7} \\
\textit{V.A. Trapeznikov Institute of Control Sciences of RAS}\\
Moscow, Russia\\
aiglush@ipu.ru}
\and
\IEEEauthorblockN{2\textsuperscript{nd} Konstantin Lastochkin}
\IEEEauthorblockA{\textit{Laboratory No. 7} \\
\textit{V.A. Trapeznikov Institute of Control Sciences of RAS}\\
Moscow, Russia\\
lastconst@ipu.ru}
}

\maketitle

\begin{abstract}
An adaptive state observer is proposed for a class of overparametrized uncertain linear time-invariant systems without restrictive requirement of their representation in the observer canonical form. It evolves the method of generalized parameters estimation-based observer design and, therefore, (\emph{i}) does not require to identify Luenberger correction gain parameters, (\emph{ii}) forms states using algebraic rather than differential equation. Additionally, the developed observer is applicable to systems with unknown output matrix and ensures exponential convergence of unmeasured state observation error under weak requirement of the regressor finite excitation. The effectiveness of the proposed solution is supported by simulation results.
\end{abstract}


\section{Introduction}
Adaptive observers \cite{b1, b2, b3, b4} of linear single-input single-output systems are capable of reconstruction of unmeasured $\xi \left( t \right) \in {\mathbb{R}^n}$ state vector and unknown parameters $\eta$ simultaneously using measured control $u\left( t \right) \in \mathbb{R}$ and output $y\left( t \right) \in \mathbb{R}$ signals. The classic adaptive observers are represented as a differential equation, which is a copy of the system model up to substitution of ideal parameters $\eta$ with their dynamic estimate $\hat \eta \left( t \right)$. Adaptive laws are usually derived on the basis of estimation-based design (Luenberger Adaptive Observer \cite{b1}, Kreisselmeier Observer \cite{b2}) or using SPR-Lyapunov design technique (Adaptive Observers with Auxiliary Input \cite{b1, b3, b4}). In the second case the differential equation of the observer additionally includes Auxiliary Input, which is necessary to ensure \emph{bona fide} fulfillment of the SPR condition for the transfer function from the parametrized parametric uncertainty to output observation error $\tilde y\left( t \right) = \hat y\left( t \right) - y\left( t \right)$. The advantage of the estimation-based design in comparison with the Lyapunov one is the freedom in the choice of the adaptive law.

Regardless of the chosen adaptive law, the above-considered observers require the system model to be represented in the observer canonical state-space form and, consequently, allows one to: \emph{i}) identify the parameters of the transfer function from $u\left( t \right)$ to $y\left( t \right)$, \linebreak \emph{ii}) reconstruct virtual state vector  $\xi \left( t \right)$ of such form. However, as far as practical scenarios are concerned, the states $x\left( t \right)$ of arbitrary (not necessarily canonical) state-space representation of the system model is usually required to be reconstructed instead. As it is demonstrated in \cite{b4}, for completely observable systems there exists nonsingular linear transformation $\xi \left( t \right) = Tx\left( t \right)$, such that the system can be rewritten in the observer canonical form. Unlike \cite{b1, b2, b3}, the observer proposed in \cite{b4} for one class of linear systems in addition to estimates $\hat \xi \left( t \right){\rm{,\;}}\hat \eta \left( t \right)$ allows one to obtain ${\hat T^{ - 1}}\left( t \right){\rm{:}} = {\hat T^{ - 1}}\left( {\hat \eta \left( t \right)} \right)$ and reconstruct original (physical) state vector $x\left( t \right) = {T^{ - 1}}\xi \left( t \right)$ using $\hat \eta \left( t \right)$. However, the estimate $\hat T\left( t \right)$ can become discontinuous because the mapping from $\eta$ to $T$ is usually nonunique (see Section VIII in \cite{b4}). In more recent papers \cite{b5, b6, b7} and many others devoted to the development of adaptive observer design methods (and even in the seminal book \cite{b1}), to the best of the authors’ knowledge, the problem of estimation of both physical states $x\left( t \right)$ and linear similarity matrix $T$ using adaptive observers has not been addressed anymore. Most of the modern studies on adaptive observers for linear systems, e.g. \cite{b5, b8}, are devoted to relaxation of the regressor persistent excitation requirement, which, considering the baseline adaptive observers, is necessary to ensure exponential convergence of the unmeasured state observation error.

In the recent paper \cite{b9} a new method of adaptive observers design has been proposed that allows one to reconstruct the unmeasured state $x\left( t \right)$ of linear systems with the matrices $A\left( \theta  \right) \in {\mathbb{R}^{n \times n}}{\rm{,\;}}B\left( \theta  \right) \in {\mathbb{R}^n}$, which are overparametrized with respect to the parameters $\theta  \in {\mathbb{R}^{{n_\theta }}}$. In \cite{b10} such approach is extended to a class of linear systems with bounded external disturbances produced by a known exosystem with unknown initial conditions. The solutions \cite{b9, b10} are based on a new method \cite{b11} of unknown parameters identification for linear regression equations (LRE) with overparametrization. Following \cite{b9, b10}, the estimate of the unmeasured state is obtained with the help of a differential equation representing an adaptive version of the Luenberger observer with adjustable parameters $\hat A\left( t \right){\rm{,\;}}\hat B\left( t \right)$ and Luenberger correction gain $\hat L\left( t \right)$. Based on the method from \cite{b11}, a special parametrization to reconstruct the above-mentioned estimates 
\newgeometry{left=1.91cm,right=1.91cm,top=1.91cm,bottom=2cm} \hspace{-15pt}
using measurable signals is proposed, which allows one to, firstly, obtain linear regression equations with respect to the parameters $\eta \left( \theta  \right)$, and then – the regression equations with respect to $A\left( \theta  \right){\rm{,\;}}B\left( \theta  \right){\rm{,\;}}L\left( \theta  \right)$. If the regressor, which depends on the input/output data is finitely exciting, then the observers from \cite{b9, b10} ensure exponential convergence of the unmeasured state observation error. The shortcomings of the solutions from \cite{b9, b10} are summarized as follows:
\begin{enumerate}
    \item [\textbf{S1.}]  a complex parametrization on the basis of the results of generalized pole placement control theory \cite{b12} is required to obtain the regression equation with respect to $L\left( \theta  \right)$;
    \item [\textbf{S2.}]  the peaking phenomenon in the course of $\hat x\left( t \right)$ transient has occurred (see Fig.1 from \cite{b9}) because the Luenberger corrective feedback is used \cite{b13}.
\end{enumerate}

The main aim of this study is to design an observer that does not require to identify $L\left( \theta  \right)$, but allows one to reconstruct unmeasured states of the system represented in an arbitrary state-space form, which does not necessarily coincide with the observer canonical one. To achieve this goal, the method is developed on the basis of the generalized parameter estimation-based observer (GPEBO) \cite{b8} and the approach to identify parameters of the linear regression equations with overparameterization \cite{b11}. Using the results from \cite{b8}, the estimate of $\xi \left( t \right)$ is obtained, then, on the basis of \cite{b9, b11} a linear regression equation with respect to ${T^{ - 1}}\left( \theta  \right)$ is parameterized to identify the parameters of this matrix instead of their direct calculation.

In general, the main salient features that distinguishes the proposed approach from previous results \cite{b1, b2, b3, b4, b5, b6, b7, b8, b9, b10} are that:
\begin{enumerate}
    \item [\textbf{F1.}]  Unlike in \cite{b1, b2, b3, b4, b5, b6, b7, b8}, the state vector is reconstructed for plant with parametric uncertainty, which is not necessarily represented in the observer canonical form;
    \item [\textbf{F2.}]  In contrast to \cite{b9, b10}, the proposed observer: (\emph{i}) does not require Luenberger correction gain identification, (\emph{ii}) is applicable to systems with unknown output matrix, (\emph{iii}) forms unmeasured state estimates using an algebraic rather than a differential equation.
    \item [\textbf{F3.}] Exponential convergence of the unmeasured states observation error is guaranteed if the regressor finite excitation requirement is met.
\end{enumerate}

Definitions of the heterogeneous mapping and the regressor finite excitation condition are used throughout this paper.

{\it \bf Definition 1.} \emph{A mapping ${\cal F}\left( x \right){\rm{:\;}}{\mathbb{R}^{{n_x}}} \to {\mathbb{R}^{{n_{\cal F}} \times {m_{\cal F}}}}$ is heterogeneous of degree ${\ell _{\cal F}} \ge 1$ if there exists ${\Xi _{\cal F}}\left( {\omega \left( t \right)} \right)=\linebreak = {\overline \Xi _{\cal F}}\left( {\omega \left( t \right)} \right)\omega \left( t \right) \in {\mathbb{R}^{{\Delta _{\cal F}} \times {n_x}}}{\rm{,\;}}{\Pi _{\cal F}}\left( {\omega \left( t \right)} \right) \in {\mathbb{R}^{{n_{\cal F}} \times {n_{\cal F}}}}$, and a mapping ${{\cal T}_{\cal F}}\left( {{\Xi _{\cal F}}\left( {\omega \left( t \right)} \right)x} \right){\rm{:\;}}{\mathbb{R}^{{\Delta _{\cal F}}}} \to {\mathbb{R}^{{n_{\cal F}} \times {m_{\cal F}}}}$ such that for all $\left| {\omega \left( t \right)} \right| > 0$ and $x \in {\mathbb{R}^{{n_x}}}$ the following conditions hold:}
\begin{equation}\label{eq1}
\begin{array}{c}
{\Pi _{\cal F}}\left( {\omega \left( t \right)} \right){\cal F}\left( x \right) = {{\cal T}_{\cal F}}\left( {{\Xi _{\cal F}}\left( {\omega \left( t \right)} \right)x} \right){\rm{, }}\\
{\rm{det}}\left\{ {{\Pi _{\cal F}}\left( {\omega \left( t \right)} \right)} \right\} \ge {\omega ^{{\ell _{_{\cal F}}}}}\left( t \right)\!{\rm{,}}{\Xi _{\cal F}}_{ij}\!\left( {\omega \left( t \right)} \right) = {c_{ij}}{\omega ^\ell }\left( t \right)\!{\rm{,}}\\
{c_{ij}} \in \left\{ {0,{\rm{ 1}}} \right\}{\rm{,\;}}\ell  > 0.
\end{array}
\end{equation}

For instance, the mapping ${\cal F}\left( x \right) = {\rm{col}}\left\{ {{x_1}{x_2}{\rm{,\;}}{x_1}} \right\}$ with ${\Pi _{\cal F}}\left( \omega  \right) = {\rm{diag}}\left\{ {{\omega ^2}{\rm{,\;}}\omega } \right\}{\rm{,\;}}{\Xi _{\cal F}}\left( \omega  \right) = {\rm{diag}}\left\{ {\omega {\rm{,\;}}\omega } \right\}$ is heterogeneous of degree ${\ell _{\cal F}} = 3.$

{\it \bf Definition 2.} \emph{The regressor $\varphi \left( t \right) \in {\mathbb{R}^n}$ is finitely exciting $\varphi \left( t \right) \in {\rm{FE}}$  over the time range $\left[ {t_r^ + {\rm{;\;}}{t_e}} \right]$  if there exists $t_r^ +  \ge 0$, ${t_e} > t_r^ +$ and $\alpha$ such that the following inequality holds:}
\begin{equation}\label{eq2}
\int\limits_{t_r^ + }^{{t_e}} {\varphi \left( \tau  \right){\varphi ^{\rm{T}}}\left( \tau  \right)d} \tau  \ge \alpha {I_n}{\rm{,}}
\end{equation}
\emph{where $\alpha > 0$ is the excitation level, $I_{n}$ is an identity matrix.}

\section{Problem Statement}

A class of linear time-invariant uncertain systems with overparametrization is considered:
\begin{equation}\label{eq3}
\begin{array}{l}
\dot x\left( t \right) = A\left( \theta  \right)x\left( t \right) + B\left( \theta  \right)u\left( t \right){\rm{,}}\\
y\left( t \right) = {C^{\rm{T}}}x\left( t \right){\rm{,\;}}x\left( {{t_0}} \right) = {x_0}{\rm{,}}
\end{array}
\end{equation}
where $x\left( t \right) \in {\mathbb{R}^n}$ is the system state with unknown initial conditions ${x_0}$, $A{\rm{:\;}}{\mathbb{R}^{{n_\theta }}} \!\!\to\!\! {\mathbb{R}^{n \times n}}{\rm{,\;}}B{\rm{:\;}}{\mathbb{R}^{{n_\theta }}} \!\!\to\!\! {\mathbb{R}^n}$ stand for known mappings with unknown parameters $\theta  \in {\mathbb{R}^{{n_\theta }}}$, $C \in {\mathbb{R}^{{n}}}$ is unknown vector\footnote{Dependencies from $\theta$ and $t$ can be further suppressed for the sake of brevity.}.. The pair $\left( {{C^{\rm{T}}}{\rm{,\;}}A\left( \theta  \right)} \right)$ is completely observable and the control $u\left( t \right) \in \mathbb{R}$ and output $y\left( t \right) \in \mathbb{R}$ signals are measurable. The following assumption related to the control signal is also adopted.

{\it \bf Assumption 1.} \emph{The control signal $u\left( t \right)$ is chosen so that for $t \ge {t_0}$ it ensures existence and boundedness of all trajectories of the system \eqref{eq3}.}

The goal is to obtain estimate $\hat x\left( t \right)$ such that:
\begin{equation}\label{eq4}
\mathop {{\rm{lim}}}\limits_{t \to \infty } \left\| {\tilde x\left( t \right)} \right\| = 0{\rm{\;}}\left( {exp } \right){\rm{, }}
\end{equation}
where $\tilde x\left( t \right) = \hat x\left( t \right) - x\left( t \right)$ denotes the state observation error.

\section{Main Result}

In accordance with the results from \cite{b4, b14}, for any completely observable linear system \eqref{eq3} there exist below-given matrices:
\begin{gather*}
{\small{
\begin{array}{c}
{T_{I}}\left( \theta  \right) = {\begin{bmatrix}
{{A^{n - 1}}\left( \theta  \right){{\cal O}_n}\left( \theta  \right)}&{{A^{n - 2}}\left( \theta  \right){{\cal O}_n}\left( \theta  \right)}& \cdots &{{{\cal O}_n}\left( \theta  \right)}
\end{bmatrix}}{\rm{,}}\\
{{\cal O}_n}\left( \theta  \right) = {\cal O}\left( \theta  \right){{\begin{bmatrix}
{{0_{1 \times \left( {n - 1} \right)}}}&1
\end{bmatrix}}^{\rm{T}}}{\rm{,}}\\
{{\cal O}^{ - 1}}\left( \theta  \right) = {{\begin{bmatrix}
{C\left( \theta  \right)}&{{{\left( {A\left( \theta  \right)} \right)}^{\rm{T}}}C\left( \theta  \right)}& \cdots &{{{\left( {{A^{n - 1}}\left( \theta  \right)} \right)}^{\rm{T}}}C\left( \theta  \right)}
\end{bmatrix}}^{\rm{T}}}{\rm{,}}
\end{array}}}
\end{gather*}
which define the similarity transformation $\xi \left( t \right) = T\left( \theta  \right)x\left( t \right)$ to rewrite the system \eqref{eq3} in the observer canonical form:
\begin{equation}\label{eq6}
\dot \xi \left( t \right)\! =\! {A_0}\xi \left( t \right) + {\psi _a}\left( \theta  \right)y\left( t \right) + {\psi _b}\left( \theta  \right)u\left( t \right){\rm{,}}
\end{equation}
\begin{equation}\label{eq7}
y\left( t \right) = C_0^{\rm{T}}\xi \left( t \right){\rm{,\;}}\xi \left( {{t_0}} \right) = {\xi _0}\left( \theta  \right) = T^{-1}\left( \theta  \right){x_0}{\rm{,}}
\end{equation}
where
\begin{gather*}
\begin{array}{c}
{\psi _a}\left( \theta  \right) = T\left( \theta  \right)A\left( \theta  \right){T^{ - 1}}\left( \theta  \right){C_0}{\rm{,\;}}{\psi _b}\left( \theta  \right) = T\left( \theta  \right)B\left( \theta  \right){\rm{,}}\\
{A_0} = {\begin{bmatrix}
{{0_n}}&{\begin{array}{*{20}{c}}
{{I_{n - 1}}}\\
{{0_{1 \times \left( {n - 1} \right)}}}
\end{array}}
\end{bmatrix}}{\rm{,\;}}\begin{array}{*{20}{c}}
{C_0^{\rm{T}} = {C^{\rm{T}}}\left( \theta  \right){T^{ - 1}}\left( \theta  \right) = }\\
{ = C\left[ {\begin{array}{*{20}{c}}
1&{0_{n - 1}^{\rm{T}}}
\end{array}} \right]}
\end{array}{\rm{,}}
\end{array}
\end{gather*}
${T_{I}}\left( \theta  \right){\rm{:}}=T^{-1}\left( \theta  \right)$, ${{\cal O}_n}$ is the $n^{th}$ column of the matrix that is inverse to ${{\cal O}^{ - 1}}\left( \theta  \right)$, $\xi \left( t \right) \in {\mathbb{R}^n}$ denotes state vector of the observer canonical form with unknown initial conditions ${\xi _0}$, the vector ${C_0} \in {\mathbb{R}^n}$ and mappings ${\psi_a}{\rm{,\;}}{\psi _b}{\rm{:\;}}{\mathbb{R}^{{n_\theta }}} \to {\mathbb{R}^n}$ are known.

Using the representation \eqref{eq6}, \eqref{eq7} of the system \eqref{eq3}, the stated goal \eqref{eq4} is reduced to the problem of identification of state $\xi \left( t \right)$ and similarity matrix ${T_{I}}\left( \theta  \right)$. In accordance with the results from \cite{b8, b9}, the following parametrizations hold for unknown parameters $\eta \left( \theta  \right) = {\rm{col}}\left\{ {{\psi _a}\left( \theta  \right){\rm{,\;}}{\psi _b}\left( \theta  \right){\rm{,\;}}{\xi _0}\left( \theta  \right)} \right\}$ and state $\xi \left( t \right)$.

{\bf{Lemma.}} \emph{The unknown parameters $\eta \left( \theta  \right)$ and unmeasured state $\xi \left( t \right)$ satisfy the following linear regression models:}
\begin{equation}\label{eq8}
\begin{array}{c}
{\cal Y}\left( t \right) = \Delta \left( t \right)\eta \left( \theta  \right)\\
\xi \left( t \right) = z\left( t \right) + {H^{\rm{T}}}\left( t \right)\eta \left( \theta  \right){\rm{,}}\\
\end{array}
\end{equation}
\vspace{-15pt}
\begin{gather*}
\begin{array}{c}
    {H^{\rm{T}}}\left( t \right) =  {\begin{bmatrix}
{\Omega \left( t \right)}&{P\left( t \right)}&{\Phi \left( t \right)}
\end{bmatrix}}{\rm{, }}\\
{\cal Y}\left( t \right) = k\left( t \right) \cdot {\rm{adj}}\left\{ {\overline \varphi \left( t \right)} \right\}\overline q\left( t \right){\rm{, }}\Delta \left( t \right) = k\left( t \right) \cdot {\rm{det}}\left\{ {\overline \varphi \left( t \right)} \right\}{\rm{,}}
\end{array}
\end{gather*}
\begin{equation}\label{eq9}
\begin{array}{c}
\dot {\overline {q}}\left( t \right) = {e^{ - \sigma \left( {t - {t_0}} \right)}}\varphi \left( t \right)q\left( t \right){\rm{,\;}}\overline q\left( {{t_0}} \right) = {0_{3n}},\\
\dot {\overline {\varphi}} \left( t \right) = {e^{ - \sigma \left( {t - {t_0}} \right)}}\varphi \left( t \right){\varphi ^{\rm{T}}}\left( t \right){\rm{,\;}}\overline \varphi \left( {{t_0}} \right) = {0_{3n \times 3n}},\\
q\left( t \right) = y - C_0^{\rm{T}}z{\rm{,\;}}\varphi \left( t \right) = {\begin{bmatrix}
{{\Omega ^{\rm{T}}}{C_0}}\\
{{P^{\rm{T}}}{C_0}}\\
{ - \Phi \left( t \right){C_0}}
\end{bmatrix}}{\rm{,}}
\end{array}
\end{equation}
\begin{equation}\label{eq10}
\begin{array}{l}
\dot z\left( t \right) = {A_K}z\left( t \right) + Ky\left( t \right){\rm{,\;}}z\left( {{t_0}} \right) = {0_n}{\rm{,}}\\
\dot \Omega \left( t \right) = {A_K}\Omega \left( t \right) + {I_n}y\left( t \right){\rm{,\;}}\Omega \left( {{t_0}} \right) = {0_{n \times n}}{\rm{,}}\\
\dot P\left( t \right) = {A_K}P\left( t \right) + {I_n}u\left( t \right){\rm{,\;}}P\left( {{t_0}} \right) = {0_{n \times n}}{\rm{,}}\\
\dot \Phi \left( t \right) = {A_K}\Phi \left( t \right){\rm{,\;}}\Phi \left( {{t_0}} \right) = {I_{n \times n}}{\rm{,}}
\end{array}
\end{equation}
\emph{and $\forall t \ge {t_e}{\rm{\;\;}}\Delta \left( t \right) \ge {\Delta _{\min }} > 0$ if $\varphi \left( t \right) \in {\rm{FE}}$; $\sigma  > 0$ is a damping ratio, $k\left( t \right) > 0$ is an amplitude modulator, ${A_K} = {A_0} - KC_0^{\rm{T}}$ is a stable matrix.}

\emph{Proof of Lemma is given in \cite{b8, b9}.}

Therefore, in accordance with the parametrization \eqref{eq8}, to estimate the state $x\left( t \right)$  we first need to obtain the estimates of $\eta$ and ${T_{I}}$. As for $\eta$, this can be done using the regression equation for ${\cal Y}\left( t \right)$. To obtain the estimate of ${T_{I}}$, first of all, it is necessary to parameterize the corresponding linear regression equation. As $\eta$  depends on $\theta$ in a known manner, then, using the results from \cite{b11}, a linear regression equation with respect to $\theta$ can be obtained, which, in turn, makes it possible to derive a regression equation with respect to ${T_{I}}$. On the basis of parameterization \eqref{eq8}, the parameters $\theta$ can be identified iff the following condition is met:
\begin{equation}\label{eq11}
\begin{array}{c}
{{\rm{det}} ^2}\left\{ {{\nabla _\theta }{\psi _{ab}}\left( \theta  \right)} \right\} > 0,\\{\psi _{ab}}\left( \theta  \right) = {{\cal L}_{ab}}\eta \left( \theta  \right){\rm{:\;}}{\mathbb{R}^{{n_\theta }}} \to {\mathbb{R}^{{n_\theta }}}{\rm{,}}
\end{array}
\end{equation}
which means that there exists an inverse transform \linebreak $\theta  = {\cal F}\left( {{\psi _{ab}}} \right){\rm{:\;}}{\mathbb{R}^{{n_\theta }}} \to {\mathbb{R}^{{n_\theta }}}$.

The mapping ${\cal F}\left( {{\psi _{ab}}} \right)$ may include singularity burden operations (for example, division by elements of the vector ${\psi _{ab}}$), thus, in addition to \eqref{eq11}, a hypothesis that the mapping ${\cal F}\left( {{\psi _{ab}}} \right)$ can be transformed into linear regression equation with respect to $\theta$ is adopted from \cite{b9}.

{\bf{Hypothesis 1.}} \emph{There exist heterogeneous in the sense of \eqref{eq1} mappings ${\cal G}\left( {{\psi _{ab}}} \right){\rm{:\;}}{\mathbb{R}^{{n_\theta }}} \to {\mathbb{R}^{{n_\theta } \times {n_\theta }}}$, ${\cal S}\left( {{\psi _{ab}}} \right){\rm{:\;}}{\mathbb{R}^{{n_\theta }}} \to {\mathbb{R}^{{n_\theta }}}$  such that:}
\begin{equation}\label{eq12}
\begin{array}{c}
{\cal S}\left( {{\psi _{ab}}} \right) = {\cal G}\left( {{\psi _{ab}}} \right){\cal F}\left( {{\psi _{ab}}} \right) = {\cal G}\left( {{\psi _{ab}}} \right)\theta {\rm{,}}\\
\end{array}
\end{equation}
\begin{gather*}
    {\Pi _\theta }\left( {\Delta \left( t \right)} \right){\cal G}\left( {{\psi _{ab}}} \right) = {{\cal T}_{\cal G}}\left( {{\Xi _{\cal G}}\left( {\Delta \left( t \right)} \right){\psi _{ab}}} \right){\rm{:\;}}{\mathbb{R}^{{\Delta _{\cal G}}}} \to {\mathbb{R}^{{n_\theta } \times {n_\theta }}}{\rm{,}}\\
{\Pi _\theta }\left( {\Delta \left( t \right)} \right){\cal S}\left( {{\psi _{ab}}} \right) = {{\cal T}_{\cal S}}\left( {{\Xi _{\cal S}}\left( {\Delta \left( t \right)} \right){\psi _{ab}}} \right){\rm{:\;}}{\mathbb{R}^{{\Delta _{\cal S}}}} \to {\mathbb{R}^{{n_\theta }}}{\rm{,}}
\end{gather*}
\emph{where $\det \left\{ {{\Pi _\theta }\left( {\Delta \left( t \right)} \right)} \right\} \ge {\Delta ^{{\ell _\theta }}}\left( t \right){\rm{,\;}}rank\left\{ {{\cal G}\left( {{\psi _{ab}}} \right)} \right\} = \linebreak = {n_\theta }{\rm{,\;}}{\ell _\theta } \ge 1$, ${\Xi _{\cal G}}\left( {\Delta \left( t \right)} \right) \in {\mathbb{R}^{{\Delta _{\cal G}} \times {n_\theta }}}$, ${\Xi _{\cal S}}\left( {\Delta \left( t \right)} \right) \in {\mathbb{R}^{{\Delta _{\cal S}} \times {n_\theta }}}$, and all mappings are known.}

Owing to the mapping ${\cal G}\left( {{\psi _{ab}}} \right)$, the parametrization \eqref{eq12} allows one to avoid possible division by combinations of elements of ${\psi _{ab}}$ in the mapping ${\cal F}\left( {{\psi _{ab}}} \right)$ and, in addition, using the property ${\Xi _{\left( . \right)}}\left( {\Delta \left( t \right)} \right) = {\overline \Xi _{\left( . \right)}}\left( {\Delta \left( t \right)} \right)\Delta \left( t \right)$, makes it possible to convert the regression equation ${\cal Y}\left( t \right) = \linebreak = \Delta \left( t \right)\eta \left( \theta  \right)$ into a measurable linear regression equation with respect to the unknown parameters $\theta$:
\begin{equation}\label{eq13}
\begin{array}{c}
{{\cal Y}_\theta }\left( t \right) = {{\cal M}_\theta }\left( t \right)\theta ,\\
\end{array}
\end{equation}
\vspace{-20pt}
\begin{gather*}
    {{\cal Y}_\theta }\left( t \right) = {\rm{adj}}\left\{ {{{\cal T}_{\cal G}}\left( {{{\overline \Xi }_{\cal G}}\left( \Delta  \right){{\cal Y}_{ab}}} \right)} \right\}{{\cal T}_{\cal S}}\left( {{{\overline \Xi }_{\cal S}}\left( \Delta  \right){{\cal Y}_{ab}}} \right){\rm{,}}\\
{{\cal M}_\theta }\left( t \right) = {\rm{det}}\left\{ {{{\cal T}_{\cal G}}\left( {{{\overline \Xi }_{\cal G}}\left( \Delta  \right){{\cal Y}_{ab}}} \right)} \right\}{\rm{,\;}}{{\cal Y}_{ab}}\left( t \right) = {{\cal L}_{ab}}{\cal Y}\left( t \right),
\end{gather*}
for which, if $\varphi \left( t \right) \in {\rm{FE}}$, then for all $t \ge {t_e}$ it holds that $\left| {{{\cal M}_\theta }\left( t \right)} \right| \ge \underline {{{\cal M}_\theta }}  > 0$  as:
\begin{gather*}
\begin{array}{c}
\left| {{{\cal M}_\theta }\left( t \right)} \right| = \left| {{\rm{det}}\left\{ {{{\cal T}_{\cal G}}\left( {{{\overline \Xi }_{\cal G}}\left( {\Delta \left( t \right)} \right){{\cal Y}_{ab}}\left( t \right)} \right)} \right\}} \right| = \\
 = \left| {{\rm{det}}\left\{ {{\Pi _\theta }\left( {\Delta \left( t \right)} \right)} \right\}} \right|\underbrace {\left| {{\rm{det}}\left\{ {{\cal G}\left( {{\psi _{ab}}} \right)} \right\}} \right|}_{ > 0} \ge \\
 \ge \left| {{\Delta ^{{\ell _\theta }}}\left( t \right)} \right|\left| {{\rm{det}}\left\{ {{\cal G}\left( {{\psi _{ab}}} \right)} \right\}} \right| > 0.
\end{array}
\end{gather*}

Then it is assumed that a condition, which is similar to \eqref{eq12}, is met to parametrize the equation with respect to ${T_{I}}$.

{\bf{Hypothesis 2.}} \emph{There exist heterogenous in the sense of \eqref{eq1} mappings ${\cal Q}\left( \theta  \right){\rm{:\;}}{\mathbb{R}^{{n_\theta }}} \to {\mathbb{R}^{n \times n}}{\rm{,\;}}{\cal P}\left( \theta  \right){\rm{:\;}}{\mathbb{R}^{{n_\theta }}} \to {\mathbb{R}^{n \times n}}$ such that:}
\begin{equation}\label{eq14}
{\cal Q}\left( \theta  \right) = {\cal P}\left( \theta  \right){{T_{I}}}\left( \theta  \right){\rm{,}}
\end{equation}
\vspace{-20pt}
\begin{gather*}
\begin{array}{c}
{\Pi _{{T_{I}}}}\left( {{{\cal M}_\theta }} \right){\cal P}\left( \theta  \right) = {{\cal T}_{\cal P}}\left( {{\Xi _{\cal P}}\left( {{{\cal M}_\theta }} \right)\theta } \right){\rm{:\;}}{\mathbb{R}^{{\Delta _{\cal P}}}} \to {\mathbb{R}^{n \times n}}{\rm{,}}\\
{\Pi _{{T_{I}}}}\left( {{{\cal M}_\theta }} \right){\cal Q}\left( \theta  \right) = {{\cal T}_{\cal Q}}\left( {{\Xi _{\cal Q}}\left( {{{\cal M}_\theta }} \right)\theta } \right){\rm{:\;}}{\mathbb{R}^{{\Delta _{\cal Q}}}} \to {\mathbb{R}^{n \times n}}{\rm{,}}
\end{array}
\end{gather*}
\emph{where ${\Xi _{\cal Q}}\left( {{{\cal M}_\theta }\left( t \right)} \right) \in {\mathbb{R}^{{\Delta _{\cal Q}} \times {n_\theta }}}$, ${\Xi _{\cal P}}\left( {{{\cal M}_\theta }\left( t \right)} \right) \in {\mathbb{R}^{{\Delta _{\cal P}} \times {n_\theta }}}$, $rank\left\{ {{\cal P}\left( \theta  \right)} \right\} = n{\rm{,\;}}{\ell _{{T_{I}}}} \ge 1{\rm{,\;}}\det \left\{ {{\Pi _{{{T_{I}}}}}\left( {{{\cal M}_\theta }\left( t \right)} \right)} \right\} \ge \linebreak \ge {\cal M}_\theta ^{{\ell _{{T_{I}}}}}\left( t \right)$ and all mappings are known.}

As like as ${\cal G}\left( {{\psi _{ab}}} \right)$ in \eqref{eq12}, the mapping ${\cal P}\left( \theta  \right)$ in \eqref{eq14} allows one to avoid possible division by polynomials with respect to $\theta$ in the mapping ${{T_{I}}}\left( \theta  \right)$ (${\cal P}\left( \theta  \right)$ consists of the denominators of the matrix ${{T_{I}}}$ elements). Owing to the property ${\Xi _{\left( . \right)}}\left( {{{\cal M}_\theta }\left( t \right)} \right) = {\overline \Xi _{\left( . \right)}}\left( {{{\cal M}_\theta }\left( t \right)} \right){{\cal M}_\theta }\left( t \right)$, if Hypothesis 2 is met, then the regression equation ${{\cal Y}_\theta }\left( t \right) = {{\cal M}_\theta }\left( t \right)\theta$ is converted into linear regression equation with respect to unknown parameters ${{T_{I}}}$:
\begin{equation}\label{eq15}
\begin{array}{c}
{{\cal Y}_{{{T_{I}}}}}\left( t \right) = {{\cal M}_{{T_{I}}}}\left( t \right){{T_{I}}}\left( \theta  \right),\\
\end{array}
\end{equation}
\vspace{-20pt}
\begin{gather*}
    \begin{array}{c}
{{\cal Y}_{{T_{I}}}}\left( t \right) = {\rm{adj}}\left\{ {{{\cal T}_{\cal P}}\left( {{{\overline \Xi }_{\cal P}}\left( {{{\cal M}_\theta }} \right){{\cal Y}_\theta }} \right)} \right\}{{\cal T}_{\cal Q}}\left( {{{\overline \Xi }_{\cal Q}}\left( {{{\cal M}_\theta }} \right){{\cal Y}_\theta }} \right){\rm{,}}\\
{{\cal M}_{{T_{I}}}}\left( t \right) = {\rm{det}}\left\{ {{{\cal T}_{\cal P}}\left( {{{\overline \Xi }_{\cal P}}\left( {{{\cal M}_\theta }\left( t \right)} \right){{\cal Y}_\theta }\left( t \right)} \right)} \right\}{\rm{,}}
    \end{array}
\end{gather*}
where, if $\varphi \left( t \right) \in {\rm{FE}}$, then for all $t \ge {t_e}$ it holds that $\left| {{{\cal M}_{{T_{I}}}}\left( t \right)} \right| \ge \underline {{{\cal M}_{{T_{I}}}}}  > 0$ as:
\begin{gather*}
\begin{array}{c}
\left| {{{\cal M}_{{T_{I}}}}\left( t \right)} \right| = \left| {{\rm{det}}\left\{ {{{\cal T}_{\cal P}}\left( {{{\overline \Xi }_{\cal P}}\left( {{{\cal M}_\theta }\left( t \right)} \right){{\cal Y}_\theta }\left( t \right)} \right)} \right\}} \right| = \\
 = \left| {{\rm{det}}\left\{ {{\Pi _{{T_{I}}}}\left( {{{\cal M}_\theta }\left( t \right)} \right)} \right\}} \right|\underbrace {\left| {{\rm{det}}\left\{ {{\cal P}\left( \theta  \right)} \right\}} \right|}_{ > 0} \ge \\
 \ge \left| {{\cal M}_\theta ^{{\ell _{{{T_{I}}}}}}\left( t \right)} \right|\left| {{\rm{det}}\left\{ {{\cal P}\left( \theta  \right)} \right\}} \right| > 0.
\end{array}
\end{gather*}

Having the regression equations \eqref{eq8} and \eqref{eq15} at hand, the estimate of $x\left( t \right)$ is written as follows:
\begin{equation}\label{eq16}
\begin{array}{l}
\hat x\left( t \right) = {{\hat T_{I}}}\left( t \right)\hat \xi \left( t \right){\rm{,}}\\
\hat \xi \left( t \right) = z\left( t \right) + {H^{\rm{T}}}\left( t \right)\hat \eta \left( \theta  \right){\rm{,}}\\
\end{array}
\end{equation}
\vspace{-10pt}
\begin{gather*}
\begin{array}{l}
    \dot {\hat {\eta}} \left( t \right) = \dot {\tilde {\eta}} \left( t \right) =  - {\gamma _\eta }\Delta \left( t \right)\left( {\Delta \left( t \right)\hat \eta \left( t \right) - {\cal Y}\left( t \right)} \right){\rm{,\;}}{\gamma _\eta } > 0,\\
\dot {\hat T}_{I}\left( t \right) = \dot {\tilde T}_{I}\left( t \right) = \\ = - {\gamma _{{{T_{I}}}}}{{\cal M}_{{T_{I}}}}\left( {{{\cal M}_{{T_{I}}}}{{\hat T_{I}}} - {{\cal Y}_{{T_{I}}}}} \right){\rm{,\;}}{\gamma _{{T_{I}}}} > 0.
\end{array}
\end{gather*}

The properties of the observer \eqref{eq16} from the point of view of the stated goal \eqref{eq4} are analyzed in the following theorem.

\textbf{Theorem.} \emph{Let $\varphi \left( t \right) \in {\rm{FE}}$ and Assumption 1, condition \eqref{eq11} and Hypotheses \eqref{eq12}, \eqref{eq14} be met, then the observer \eqref{eq16} ensures that both the goal \eqref{eq4} is achieved, and the following equalities hold:}
\begin{displaymath}
\mathop {{\text{lim}}}\limits_{t \to \infty } \left\| {\tilde \eta \left( t \right)} \right\| = {\text{0 }}\left( {\exp } \right){\text{, }}\mathop {{\text{lim}}}\limits_{t \to \infty } \left\| {{{\tilde {T}_{I}}}\left( t \right)} \right\| = {\text{0 }}\left( {\exp } \right).    
\end{displaymath}

\emph{Proof of Theorem is postponed to Appendix.}

Thus, observer \eqref{eq16} is based on the GPEBO technique from \cite{b8}. In contrast to the results from \cite{b8}, the original states $x\left( t \right)$ of the initial model \eqref{eq3} are reconstructed in addition to the virtual ones $\xi \left( t \right)$ of the system \eqref{eq6}, which is represented in the observer canonical form. Unlike \cite{b9, b10}, the observer \eqref{eq16}: ({\it i}) is derived without complicated parametrization of the regression equation for the Luenberger correction gain and its further identification, ({\it ii}) is based on the algebraic \eqref{eq16} rather than a differential equation to calculate estimate of $x\left( t \right)$, which helps to avoid the peaking phenomenon.

\textbf{Remark 1.} \emph{In contrast to the proof of Theorem from \cite{b9}, the proof of the above-presented one does not formally require to apply the following switching operator in the definition of the adaptive gains:}
\begin{equation}\label{eq17}
\begin{gathered}
{\gamma _{{T_{I}}}}\left( t \right){\text{:}} = \left\{ \begin{gathered}
  0{\text{, if }}\Delta \left( t \right) < \rho  \in \left[ {{\Delta _{\rm{min}}}{\text{; }}{\Delta _{\max }}} \right){\text{,}} \hfill \\
  \frac{{{\gamma _1}}}{{\mathcal{M}_{{T_{I}}}^2\left( t \right)}}{\text{ otherwise,}} \hfill \\ 
\end{gathered}  \right.{\text{, }}\\
{\gamma _\eta }\left( t \right){\text{:}} = \left\{ \begin{gathered}
  0{\text{, if }}\Delta \left( t \right) < \rho  \in \left[ {{\Delta _{\rm{min}}}{\text{; }}{\Delta _{\max }}} \right){\text{,}} \hfill \\
  \frac{{{\gamma _1}}}{{{\Delta ^2}\left( t \right)}}{\text{ otherwise,}} \hfill \\ 
\end{gathered}  \right.
\end{gathered}    
\end{equation}	
\emph{but in practice the adaptive gain should still be chosen according to \eqref{eq17} or the following conditions in order to provide the desired convergence rate of the errors} $\tilde \eta \left( t \right){\text{, }}{\tilde T_{I}}\left( t \right)$:
\begin{displaymath}
\begin{gathered}
{\gamma _\eta } \geqslant \left( {{\gamma _{{\rm{min}}}} \sim \tfrac{1}{{{\Delta ^2}\left( t \right)}}} \right) > 0{\text{, }}{\gamma _{{{T_{I}}}}} \geqslant \left( {{\gamma _{{\rm{min}}}} \sim \tfrac{1}{{\mathcal{M}_\kappa ^2\left( t \right)}}} \right) > 0.
\end{gathered}
\end{displaymath}

\section{Numerical Experiments}
The system from the experimental section of \cite{b9} has been considered:
\begin{equation}\label{eq18}
\begin{gathered}
  \dot x = {\begin{bmatrix}
  0&{{\theta _1} + {\theta _2}}&0 \\ 
  { - {\theta _2}}&0&{{\theta _2}} \\ 
  0&{ - {\theta _3}}&0 
\end{bmatrix}} x + {\begin{bmatrix}
  0 \\ 
  0 \\ 
  {{\theta _3}} 
\end{bmatrix}} u{\text{,}} \hfill \\
  y = {\begin{bmatrix}
  0&0&1 
\end{bmatrix}} x. \hfill \\ 
\end{gathered}
\end{equation}

Being transformed into the observer canonical form \eqref{eq6}, the system \eqref{eq18} was described by the following vectors:
\begin{displaymath}
{\psi _a} = {\begin{bmatrix}
  0 \\ 
  { - \left( {{\theta _1} + {\theta _2} + {\theta _3}} \right){\theta _2}} \\ 
  0 
\end{bmatrix}} {\text{,\;}}{\psi _b} = {\begin{bmatrix}
  {{\theta _3}} \\ 
  0 \\ 
  {{\theta _3}{\theta _2}\left( {{\theta _2} + {\theta _1}} \right)} 
\end{bmatrix}}{\text{,}}
\end{displaymath}
where 
\begin{displaymath}
{\psi _{ab}}\left( \theta  \right) = col\left\{ { - \left( {{\theta _1} + {\theta _2} + {\theta _3}} \right){\theta _2}{\text{, }}{\theta _3}{\text{, }}{\theta _3}{\theta _2}\left( {{\theta _2} + {\theta _1}} \right)} \right\}.    
\end{displaymath}

The mappings ${\mathcal{T}_\mathcal{S}}\left( . \right){\text{, }}{\mathcal{T}_\mathcal{G}}\left( . \right)$ were implemented as in \cite{b9}:
\begin{displaymath}
\begin{gathered}
    {\mathcal{T}_\mathcal{S}}\!\left( {{{\overline \Xi }_\mathcal{S}}\left( \Delta  \right){\mathcal{Y}_{ab}}} \right) \!\!=\!\! {\begin{bmatrix}
  {{\mathcal{Y}_{2ab}}{{\left( {{\mathcal{Y}_{1ab}}{\mathcal{Y}_{2ab}} \!+\! \Delta {\mathcal{Y}_{3ab}}} \right)}^2} \!-\! \mathcal{Y}_{2ab}^4{\mathcal{Y}_{3ab}}} \\ 
  { - {\mathcal{Y}_{1ab}}{\mathcal{Y}_{2ab}} - \Delta {\mathcal{Y}_{3ab}}} \\ 
  {{\mathcal{Y}_{2ab}}{\mathcal{Y}_{1ab}}} 
\end{bmatrix}}{\text{,}}\\
{\mathcal{T}_\mathcal{G}}\left( {{{\overline \Xi }_\mathcal{G}}\left( \Delta  \right){\mathcal{Y}_{ab}}} \right) =  {\begin{bmatrix}
  {\mathcal{Y}_{2ab}^3\left( {{\mathcal{Y}_{1ab}}{\mathcal{Y}_{2ab}} + \Delta {\mathcal{Y}_{3ab}}} \right)} \\ 
  {\mathcal{Y}_{2ab}^2} \\ 
  {\Delta {\mathcal{Y}_{1ab}}} 
\end{bmatrix}}.
\end{gathered}    
\end{displaymath}

The mapping ${{T_{I}}}\left( \theta  \right)$ was written as follows:
\begin{displaymath}
{{T_{I}}}\left( \theta  \right) = {\begin{bmatrix}
  { - \tfrac{{{\theta _1} + {\theta _2}}}{{{\theta _3}}}}&0&{\tfrac{1}{{{\theta _2}{\theta _3}}}} \\ 
  0&{ - \tfrac{1}{{{\theta _3}}}}&0 \\ 
  1&0&0 
\end{bmatrix}}.    
\end{displaymath}

In their turn, the mappings $\mathcal{Q}\left( \theta  \right){\text{, }}\mathcal{P}\left( \theta  \right)$ took the form:
\begin{displaymath}
\begin{gathered}
\mathcal{Q}\left( \theta  \right) = {\begin{bmatrix}
  { - {\theta _2}\left( {{\theta _1} + {\theta _2}} \right)}&0&1 \\ 
  0&{ - 1}&0 \\ 
  1&0&0 
\end{bmatrix}}{\text{, }}\\
\mathcal{P}\left( \theta  \right) = diag\left\{ {{\theta _2}{\theta _3}{\text{, }}{\theta _3}{\text{, 1}}} \right\}.
\end{gathered}
\end{displaymath}

Therefore, the mappings ${\mathcal{T}_\mathcal{Q}}\left( . \right){\text{, }}{\mathcal{T}_\mathcal{P}}\left( . \right)$ were implemented as follows:
\begin{displaymath}
\begin{gathered}
{{\cal T}_{\cal Q}}\left( {{{\overline \Xi }_{\cal Q}}\left( {{{\cal M}_\theta }} \right){{\cal Y}_\theta }} \right) = {\begin{bmatrix}
  { - {\mathcal{Y}_{2\theta }}\left( {{\mathcal{Y}_{1\theta }} + {\mathcal{Y}_{2\theta }}} \right)}&0&{\mathcal{M}_\theta ^2} \\ 
  0&{ - {\mathcal{M}_\theta }}&0 \\ 
  {{\mathcal{M}_\theta }}&0&0 
\end{bmatrix}}{\text{, }}\\
{{{\cal T}_{\cal P}}\left( {{{\overline \Xi }_{\cal P}}\left( {{{\cal M}_\theta }} \right){{\cal Y}_\theta }} \right)} = diag\left\{ {{\mathcal{Y}_{2\theta }}{\mathcal{Y}_{3\theta }}{\text{, }}{\mathcal{Y}_{3\theta }}{\text{, }}{\mathcal{M}_\theta }} \right\}.
\end{gathered}
\end{displaymath}

The control signal was defined as a P-controller \linebreak $u =  - 25\left( {r - y} \right)$. The reference signal $r$ and parameters of the system \eqref{eq18} were chosen as:
\begin{equation}\label{eq19}
r = 100 + 2{\text{.5}}{e^{ - t}}{\text{sin}}\left( {10t} \right){\text{, }}{\theta _1} = {\theta _2} = 1{\text{, }}{\theta _3} =  - 1.    
\end{equation}

The parameters of filters \eqref{eq9}, \eqref{eq10} and adaptive laws \eqref{eq16} + \eqref{eq17} were set as:
\begin{equation}\label{eq20}
K = {{\begin{bmatrix}
  3&3&1 
\end{bmatrix}}^{\text{T}}}{\text{, }}k = {10^7}{\text{, }}\sigma  = 5{\text{, }}\rho  = 0{\text{.1, }}{\gamma _1} = 1.
\end{equation}

The parameters of the adaptive laws \eqref{eq16} were chosen so as to ensure rate of convergence, which coincided with the one for the adaptive laws in \cite{b9}.

Figures 1 and 2 depict transients of the error $\tilde x\left( t \right)$ and estimates $\hat \eta \left( t \right){\text{, }}{\hat T_{I}}\left( t \right)$.
\begin{figure}[htbp]
\centerline{\includegraphics[scale=0.63]{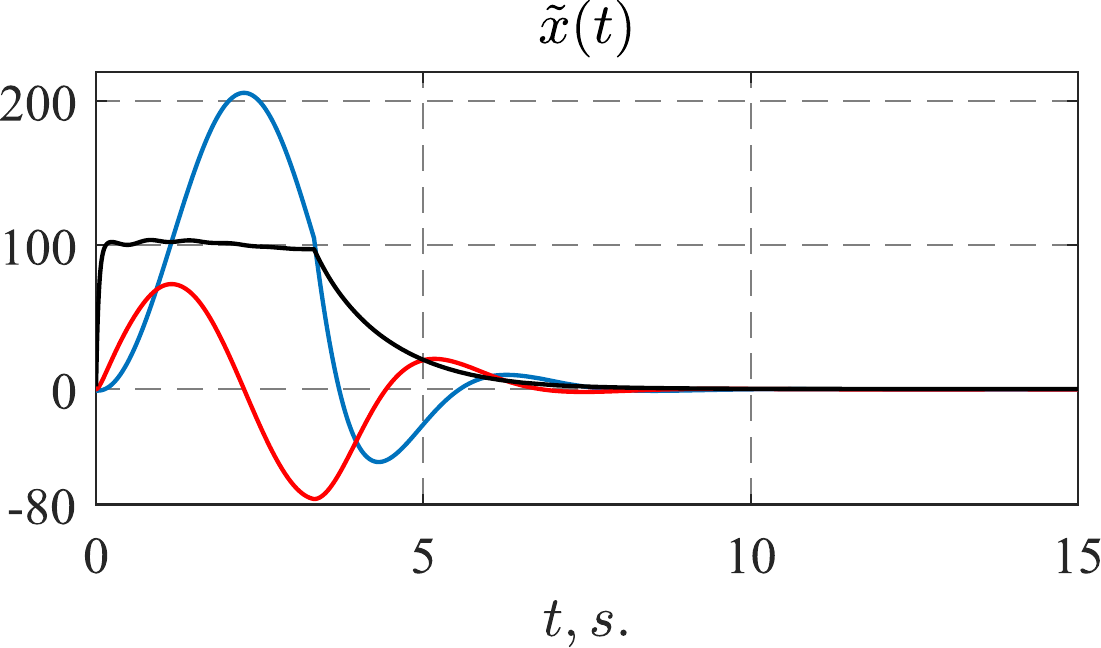}}
	\caption{Transient curves of $\tilde x\left( t \right)$.}
	\label{fig1}
\end{figure}
\begin{figure}[htbp]
\centerline{\includegraphics[scale=0.6]{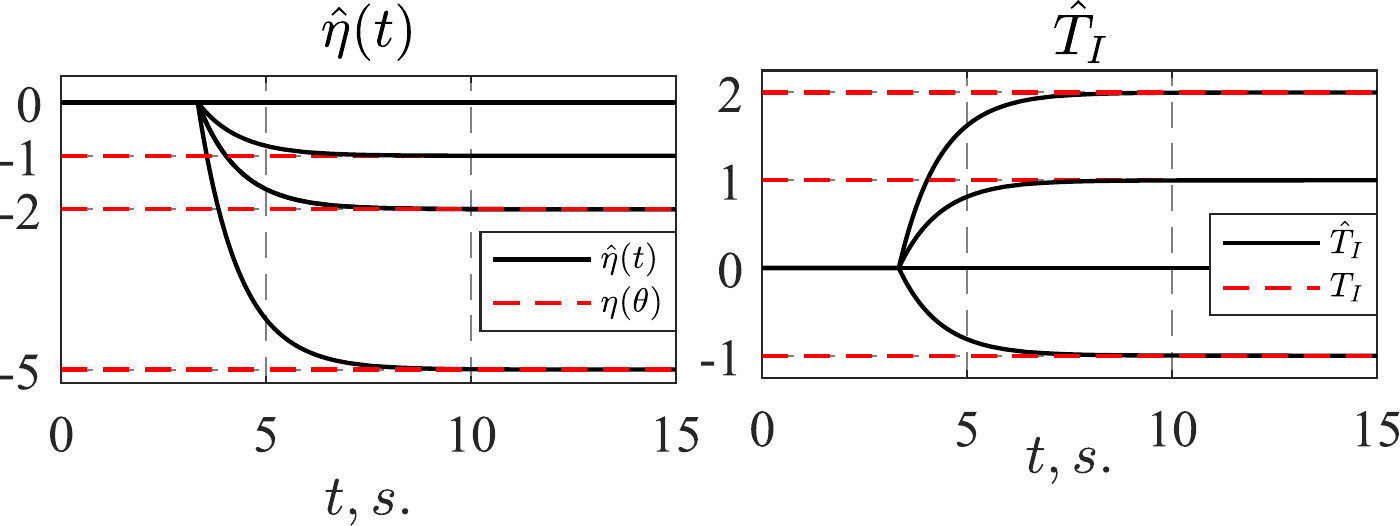}}
	\caption{Transient curves of $\hat \eta \left( t \right)$ and ${\hat T_{I}}\left( t \right)$}
	\label{fig2}
\end{figure}

The simulation results validated the conclusions of the theoretical analysis. After the condition \eqref{eq2} had been met over the time range $\left[ {0{\text{; 3}}} \right]$, the proposed adaptive observer \eqref{eq16} ensured that both observation and tracking errors converged to zero in full accordance with the goal \eqref{eq4}. Compared to the results obtained in \cite{b9}, the parameter estimation-based observer proposed in this paper allows one to avoid overshoot for the transient curves of $\tilde x\left( t \right)$ over the time range $\left[ {{\text{5; 10}}} \right]$. Such overshoot in \cite{b9} can be explained by the peaking phenomenon \cite{b13}, caused by inclusion of a linear system with nonzero initial conditions into a loop with the correction feedback gain $\hat L\left( \theta  \right)\left( {\hat y\left( t \right) - y\left( t \right)} \right)$. The observer proposed in this paper uses an algebraic equation to calculate the unmeasured state estimates, which, unlike a differential one, does not require to apply the Luenberger correction gain and helps to avoid peacking phenomenon.

\section{Conclusion}

A new adaptive observer based on the parameter estimation was proposed to reconstruct the unmeasured state of linear systems with overparameterization. In contrast to the existing solutions \cite{b1,b2,b3,b4,b5,b6,b7,b8}, such systems were represented in an arbitrary state-space form. Unlike \cite{b9,b10}, the proposed observer did not require the Luenberger gain $L\left( \theta  \right)$ identification or the vector $C$ to be {\it a priori} known, and allowed one to obtain estimate of the unmeasured state vector on the basis of an algebraic rather than a differential equation. Future research scope includes extension of the obtained results to a class of systems affected by external disturbances.

\appendix
\renewcommand{\theequation}{A\arabic{equation}}
\setcounter{equation}{0}  

\subsection{Proof of Theorem} 
The observation error $\tilde x\left( t \right) = \hat x\left( t \right) - x\left( t \right)$ is expanded as:
\begin{equation}\label{eqA1}
\begin{gathered}
  \tilde x\!\left( t \right) \!=\! {{\hat T_{I}}}\!\left( t \right)\hat \xi \!\left( t \right) \!-\! {{T_{I}}}\!\left( t \right)\xi \!\left( t \right) \pm {{\hat T_{I}}}\!\left( t \right)\xi \!\left( t \right) = \\
   \!=\! {{\hat T_{I}}}\!\left( t \right)\left( {\hat \xi \!\left( t \right) \!-\! \xi \!\left( t \right)} \right) \!+\! \left( {{{\hat T_{I}}}\!\left( t \right) \!-\! {{T_{I}}}\!\left( \theta  \right)} \right)\!\xi\! \left( t \right) \!=\! \\
   \!=\! {{\hat T_{I}}}\!\left( t \right)\left( {z\!\left( t \right) \!+\! {H^{\text{T}}}\!\left( t \right)\hat \eta \!\left( t \right) \!-\! z\!\left( t \right) \!-\! {H^{\text{T}}}\!\left( t \right)\eta \!\left( \theta  \right)} \right) \!+\!\\
   \!+\! \left( {{{\hat T_{I}}}\!\left( t \right) \!-\! {{T_{I}}}\!\left( \theta  \right)} \right)\xi \!\left( t \right) \!=\! {{\hat T_{I}}}\!\left( t \right){H^{\text{T}}}\!\left( t \right)\tilde \eta \!\left( t \right) \!+\!\\
   + {{\tilde T_{I}}}\left( t \right)\xi \left( t \right) \pm {{T_{I}}}\left( \theta  \right){H^{\text{T}}}\left( t \right)\tilde \eta \left( t \right) = \\
   = {{T_{I}}}\left( \theta  \right){H^{\text{T}}}\left( t \right)\tilde \eta \left( t \right) + {{\tilde T_{I}}}\left( t \right){H^{\text{T}}}\left( t \right)\tilde \eta \left( t \right) +\\
   + {{\tilde T_{I}}}\left( t \right)\xi \left( t \right). 
\end{gathered}  
\end{equation}

If Assumption 1 is met, then the following inequalities hold as the matrix ${A_K}$ is a Hurwitz one:
\begin{equation}\label{eqA2}
\begin{gathered}
  \left\| {{H^{\text{T}}}\left( t \right)} \right\| \leqslant {H_{{\text{max}}}}{\text{, }}\left\| {{{T_{I}}}\left( \theta  \right)} \right\| \leqslant \overline{T}_{max}{\text{, }}\\
  \left\| {\xi \left( t \right)} \right\| \leqslant {\xi _{\max }}.  
\end{gathered}
\end{equation}

Taking \eqref{eqA2} into consideration, the following upper bound is written for the error \eqref{eqA1}:
\begin{equation}\label{eqA3}
\begin{gathered}
\left\| {\tilde x\left( t \right)} \right\| \leqslant \overline{T}_{max}{H_{\max }}\left\| {\tilde \eta \left( t \right)} \right\| +\\
+ \left\| {{{\tilde T_{I}}}\left( t \right)} \right\|{H_{{\text{max}}}}\left\| {\tilde \eta \left( t \right)} \right\| + \left\| {{{\tilde T_{I}}}\left( t \right)} \right\|{\xi _{\max }}.
\end{gathered}
\end{equation}

If $\varphi \left( t \right) \in {\text{FE}}$, then, considering the parametrization \eqref{eq13}, for all $t \geqslant {t_e}$ it holds that $\Delta \left( t \right) \geqslant {\Delta _{\min }} > 0{\text{, }}\left| {{\mathcal{M}_{{T_{I}}}}\left( t \right)} \right| \geqslant \linebreak \geqslant \underline {{\mathcal{M}_{{T_{I}}}}}  > 0$, and, in their turn, the upper bounds of differential equations \eqref{eq16} for all $t \geqslant {t_e}$ are written as:
\begin{equation}\label{eqA4}
\begin{gathered}
\left\| {\tilde \eta \left( t \right)} \right\| \leqslant {e^{ - {\gamma _{_\eta }}\Delta _{\min }^2\left( {t - {t_e}} \right)}}\left\| {\tilde \eta \left( {{t_0}} \right)} \right\|{\text{, }}\\
\left\| {{{\tilde T_{I}}}\left( t \right)} \right\| \leqslant {e^{ - {\gamma _{_{{{T_{I}}}}}}\underline {\mathcal{M}_{{{T_{I}}}}^2} \left( {t - {t_e}} \right)}}\left\| {{{\tilde T_{I}}}\left( {{t_0}} \right)} \right\|{\text{,}}
\end{gathered}
\end{equation}
frow which it is concluded that the errors $\left\| {\tilde \eta \left( t \right)} \right\|{\text{, }}\left\| {{{\tilde T_{I}}}\left( t \right)} \right\|$ and $\left\| {\tilde x\left( t \right)} \right\|$ converge to zero exponentially, as was to be proved in Theorem.

\end{document}